# A Novel Interface Database of Graphene Nanoribbon from Density Functional Theory

Ao Wu[1, 2], Jiangxue Huang[1, 2], Qijun Huang[1, 2], Jin He[1, 2], Hao Wang[1, 2, *], and Sheng Chang[1, 2, *]

[1] Key Laboratory of Artificial Micro- and Nano-Structures of Ministry of Education, School of Physics and Technology, Wuhan University, Wuhan, Hubei 430072, P. R. China

[2] School of Microelectronics, Wuhan University, Wuhan, Hubei 430072, P. R. China

[*]Corresponding authors (email: wanghao@whu.edu.cn and changsheng@whu.edu.cn)

## Abstract

Interfaces play a crucial role in determining the overall performance and functionality of electronic devices and systems. Driven by the data science, machine learning (ML) reveals excellent guidance for material selection and device design, in which an advanced database is crucial for training models with state-of-the-art (SOTA) precision. However, a systematic database of interfaces is still in its infancy due to the difficulties in collecting raw data in experiment and the expensive first-principles computational cost in density functional theory (DFT). In this paper, we construct ample interface structures of graphene nanoribbons (GNR), whose interfacial morphology can be precisely fabricated based on specific molecular precursors. The GNR interfaces serve as promising candidates since their bandgaps can be modulated. Their physical properties including energy bands and density of states (DOS) maps are obtained under reasonable calculation parameters. This database can provide theoretical guidance for the design of electronic devices and accelerate the ML study of various physical quantities.

## Background and Summary

Graphene, a single layer of carbon atoms arranged in a hexagonal lattice, possesses a range of unique and remarkable properties such as exceptional electrical and thermal conductivity, mechanical strength, biocompatibility, flexibility and transparency. For semiconductor materials, a limiting fact is that the intrinsic graphene has no energy band gap and a promising option is that the transformation of graphene into a quasi-one-dimensional nanoribbon structure can open the band gap [1] [2]. Then, several modulation methods can be applied to derive diverse interface structures, including width tailoring [3], segment splicing [4], nanohole modification [5] and impurity doping [6]. These methods modulate energy band structures and offer flexibility in device design. The negative differential resistance (NDR) devices can be constructed by GNRs with different bandgaps. These devices offer higher quality due to the absence of issues like lattice mismatch, as compared to conventional interfaces [7]. The tunnel field-effect transistor (TFET) based on GNRs achieve subthreshold swing below 60 mV/decade at room temperature [8]. The cold source field effect transistors constituted by GNRs with specific edge break through the limitation of the thermal injection working mechanism in conventional metal-oxide-semiconductor FETs (MOSFET) and realize lower subthreshold swing and better switching characteristics [9].

First-principles calculation is a powerful and fundamental approach in the field of computational materials science and device physics. Unlike empirical methods that rely on experimental data, first-principles calculations can predict the properties of materials and devices even in situations where experimental data may be limited or nonexistent. This predictive power is essential for exploring novel materials and designing new devices with desired functionalities.

Here, following this paradigm, we quantitatively model crucial GNR interfaces and acquire their physical properties to constitute the database. The GNRs utilized to form interface are categorized into the armchair and the zigzag types based on the shape of their edges. Then the two selected basic structures fabricate diverse interface structures according to various coupling modes. Finally, the energy band structure and DOS are obtained by the high-throughput DFT calculations, with their projected contributions by the two components, respectively. There is now a total of 397 armchair and 345 zigzag interfaces in Table 1, and more structures are being calculated. All data and visualization codes are available on the Github. Our work provides an essential database for critical interface structure exploration.

Table 1. Structure of the entire database and statistics on the number of each item.

| Edge Type | Items | Files Count |
|---|---|---|
| Armchair (Total: 397) | CIF-After Structure Relaxation | 397 × 2 = 794 |
| | CIF- Before Structure Relaxation | 397 × 2 = 794 |
| | Projected Band Structure | 397 × 3 = 1191 |
| | Projected Density of States | 397 × 2 = 794 |
| Zigzag (Total: 345) | CIF-After Structure Relaxation | 345 × 2 = 690 |
| | CIF- Before Structure Relaxation | 345 × 2 = 690 |
| | Projected Band Structure | 345 × 3 = 1035 |
| | Projected Density of States | 345 × 2 = 690 |

## Methods

### Modeling Interfaces

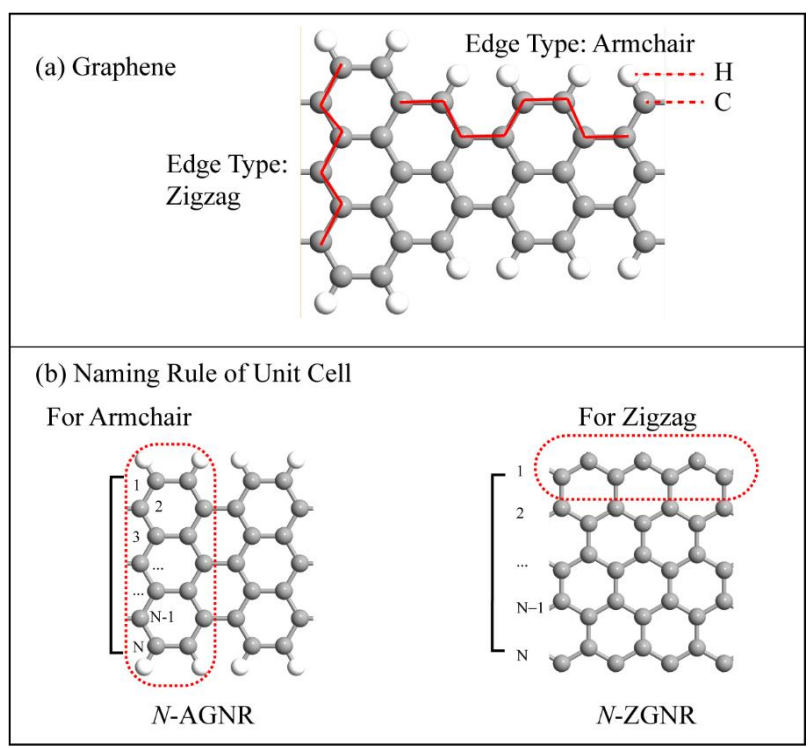

Figure 1. The diagrams of armchair GNR (AGNR) and zigzag GNR (ZGNR) and the naming rules of unit cells in this thesis.

The GNR has two types of basic components based on its edge configuration, the armchair and zigzag edges, as shown in Figure 1(a). There are some differences in the naming rules for the two types of GNRs in this thesis. The N-AGNR means the number of carbon atoms perpendicular to the armchair boundary, and the N-ZGNR means the number of zigzag edges, each of which actually contains two carbon atoms. Based on each type of edge, GNRs of different widths are spliced together to form the interface. The coupling mode at the interface plays a crucial role in the electronic state distribution of interface [10]. We combine the two GNRs in Figure 2 to create different interface structures. In the case of Figure 2(a), the naming rule for the interface formed based on the armchair segments is A-N-A-B-ID, where A stands for armchair, N stands for folds of this unit cell, A/B stands for the width of each segment, and finally the ID is the serial number of this type. Therefore, here is a variety of structures coupled by a 3-fold 7AGNR with a 15AGNR. Similarly, in Figure 2 (b), the naming rule is Z-N-A-B-ID, where Z stands for the zigzag boundary, the significance of N and ID remains unchanged, and A/B is the number of zigzag edges. Therefore, here is a variety of structures coupled by a 4-fold 6ZGNR with a 9ZGNR.

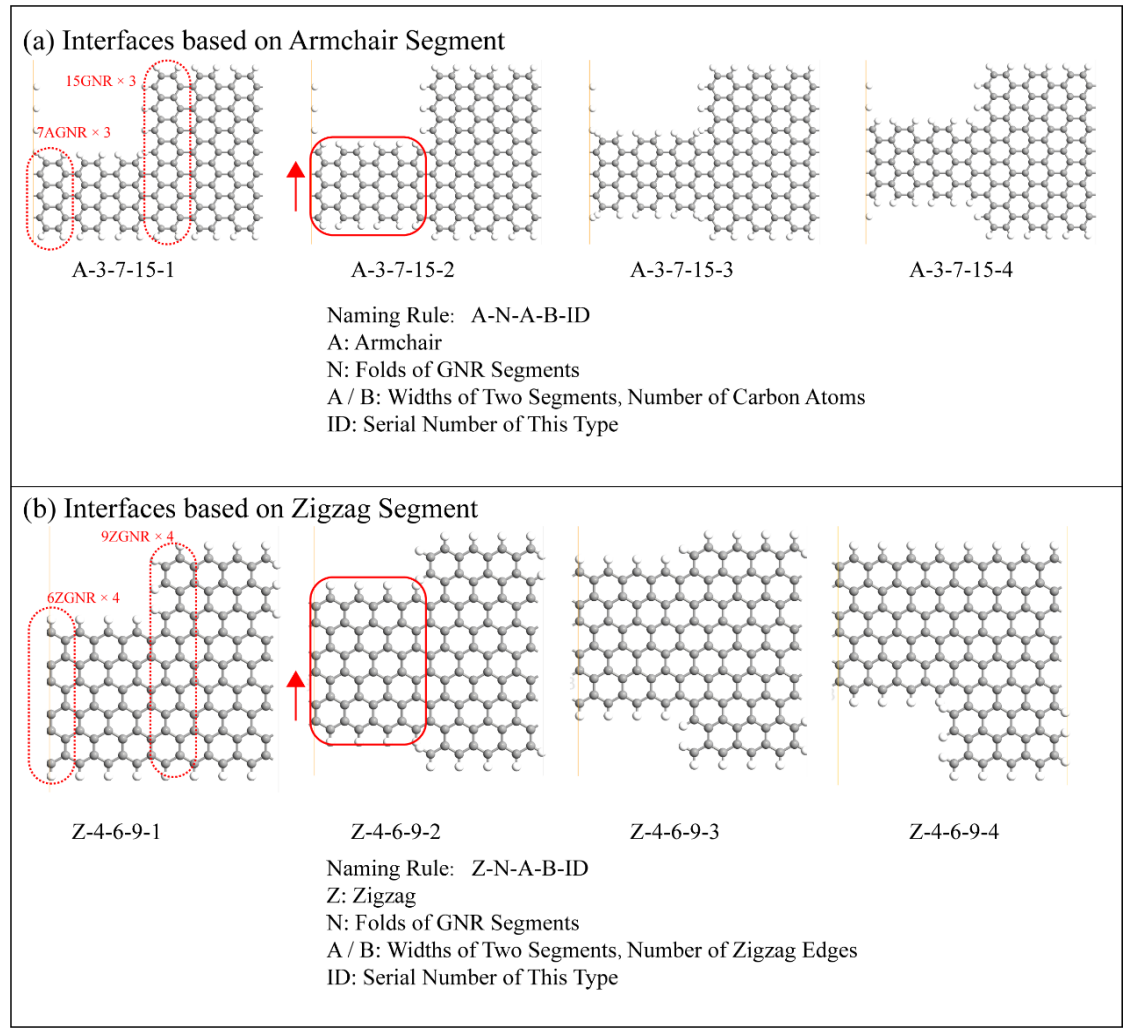

Figure 2. Examples of interface structures based on armchair and zigzag segments, respectively, and their naming rules.

**Computational Details of Density Functional Theory**

DFT is a method for studying the electronic structure of many-body system and is one of the most commonly used methods in the fields of condensed matter physics computational materials science and quantum chemistry. For a quantum many-body system, the Schrödinger equation is

$$\left[-\frac{h^2}{2m}\sum_{i=1}^{N}\nabla_i^2 + \sum_{i=1}^{N}V(r_i) + \sum_{i=1}^{N}U(r_i, r_j)\right]\psi = E\psi \tag{1}$$

where the three terms in square brackets are the kinetic energy of each electron $-\frac{h^2}{2m}\sum_{i=1}^{N}\nabla_i^2$, the energy of interaction between each electron and all nuclei $\sum_{i=1}^{N}V(r_i)$, and the energy of interaction between different electrons $\sum_{i=1}^{N}U(r_i, r_j)$. For this Hamiltonian quantity, $\psi$ is the electron wavefunction which is a function of the spatial coordinates of each of the $N$ electrons namely $\psi = \psi(r_1, ..., r_N)$, $E$ is the ground state energy of electrons. The ground state energy is independent of time, so this equation is time-neutral. However, the all-electron wavefunction has too high a dimension of $N \times 3$. The single-electron approximation needs to be introduced to allow the all-electron wavefunction to take the form of a Hartree Product $\psi = \psi(r_1)\psi(r_2)...\psi(r_N)$. The researchers measure the probability of only $N$ electrons appearing at a particular coordinate, equal to $\psi^*(r)\psi(r)$, where the asterisks are conjugate. Based on the single-electron wavefunction, the charge density at a specific location in space is written as $n(r) = 2\sum_{i=1}^{N}\psi^*(r)\psi(r)$, where 2 represents the two spins of the electron.

The Hohenberg-Kohn theorem states that the energy of the ground state obtained from the Schrödinger equation is a unique function of the charge density. The ground state charge density uniquely determines all the properties of the ground state, including the energy and the wavefunction. The ground state energy is expressed as $E[n(r)]$, where the square brackets are the functional. The charge density that minimizes the generalized function is the true charge density. To solve the true charge density is expressed as calculate a full set of equations, where each equation relates to only one electron. The Kohn-Sham equation is

$$\left[-\frac{h^2}{2m}\nabla^2 + V(r)\right]\psi_i = E\psi_i \tag{2}$$

its solution is a one-electron wavefunction $\psi_i$ that depends only on three spatial variables. The solution process is a self-consistent cycle. First, defines an initial attempted charge density $n(r)$. Second, solves the Kohn-Sham equation determined from the attempted charge density to obtain the single-electron wave function $\psi_i$. Third, solve for the charge density $n_{KS}(r)$. Fourth, compare the two charge densities and iterate the loop until the accuracy requirement is satisfied. Above is the computational flow of DFT and the database is obtained in this self-consistent computational loop.

The *ab-initio* computational software QuantumATK [11] is applied in our research. We choose the generalized gradient approximation (GGA) for the electrical properties of this material system. The Monkhorst-Pack k-meshes are set to $1 \times 1 \times 32$ for armchair edge and $1 \times 1 \times 96$ for zigzag edge and the density mesh cutoff is set to 80 Hartree. We add vacuum layers of 20 angstroms to neglect the interaction between stripes. The structural relaxations are performed in advance and allowed until the absolute value of atomic force is less than 0.05 eV/angstrom. The self-consistent calculation is limited to the tolerance of $10^{-4}$ eV.

**Data records**

As shown in Figure 3, each structure in the database contains three elements. The first is the crystal structure, which is represented by a crystallographic information file (CIF), produced by the PyMatGen [12] package. This file records information about the unit cell structure, the number of atoms and their configurations, and can be used as a database for first-principles computational software and ML algorithms to reproduce the crystal structure. We model the interface and label the two segments that make up the interface as left and right in panel 3(a). Then, based on the aforementioned computational parameters, the energy band structure and DOS are computed and the projected components of the two segments are obtained as shown in Panels 3(b) and (c), respectively.

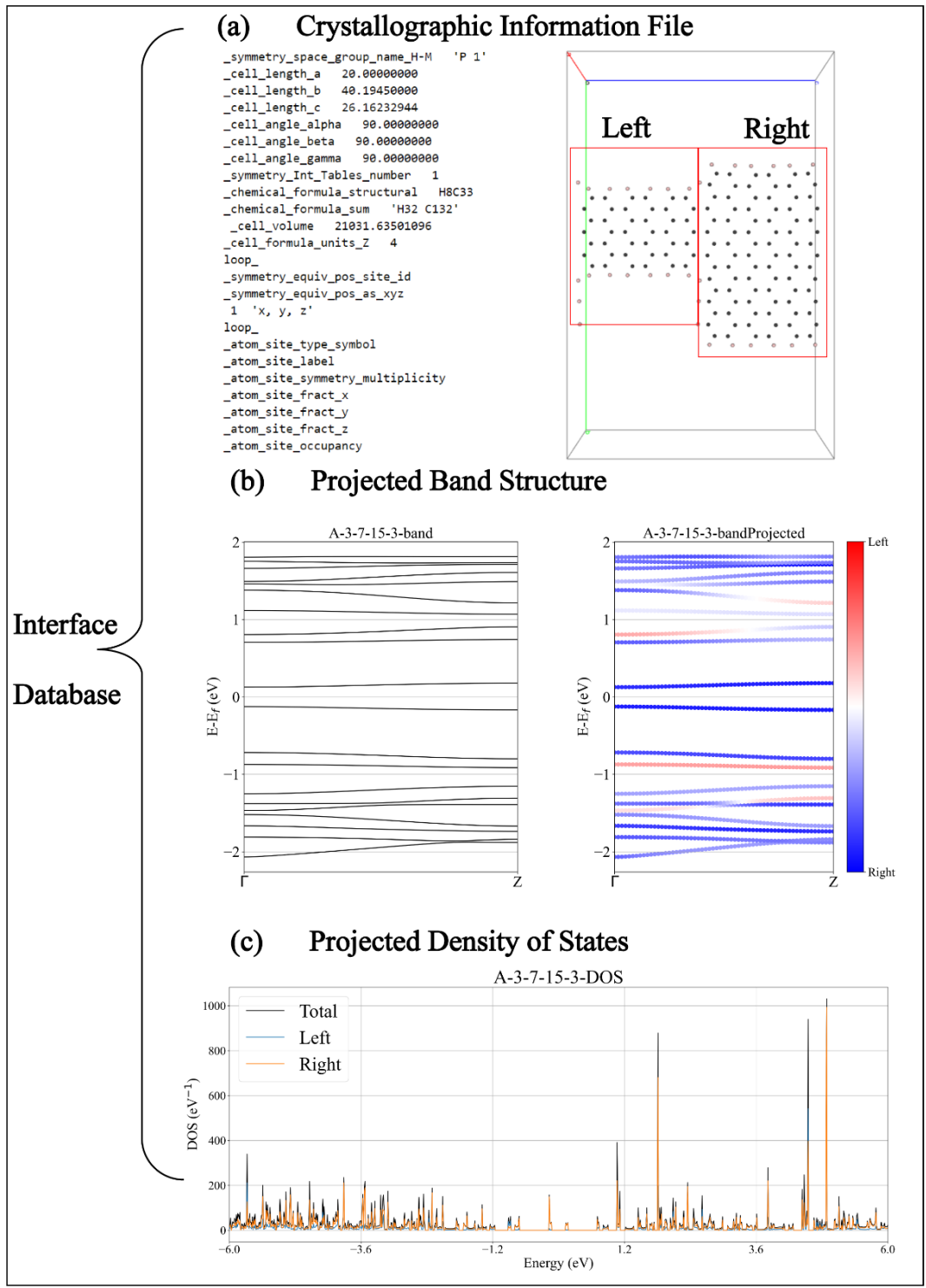

Figure 3. Framework of our interface database, divided into CIF file, projected band structure and projected density of states (PDOS).

The projected band structure is a method for visualizing the contribution of different components to the energy band, which is based on the principle of calculating the weight of each eigenvalue for a given projected space. The calculations are as follows: the eigenfunctions and eigenvalues are

obtained by the usual method, that is, solving the following eigenequations:

$$H\psi_{n,k} = E\psi_{n,k} \tag{3}$$

Next, define the projection operator $\widehat{p_M}$ on the subspace $M$, and obtain the weight $\omega_{n,k}^M$ of band and wavefunction $(E_{n,k}, \psi_{n,k})$.

$$\omega_{n,k}^M = \langle \psi_{n,k} | \widehat{p_M} | \psi_{n,k} \rangle \tag{4}$$

This weight satisfies the wave function of the normalization condition. We plot the energy band structure and projected energy band of interface A-3-7-15-3 in Figure 3(b). The proportions of the left and right segments are indicated by red and blue colors, respectively. The projected DOS (PDOS) is a way to visualize the contribution of different components to the density of states. From the definition of the DOS:

$$D(\epsilon) = \sum \delta(\epsilon - \epsilon_n) \tag{5}$$

The PDOS is obtained as:

$$D(\epsilon) = \sum \delta(\epsilon - \epsilon_n) \langle \psi_n | \widehat{p_M} | \psi_n \rangle \tag{6}$$

The above plentiful physical images serve as an excellent theoretical basis to guide interface characterization and device design.

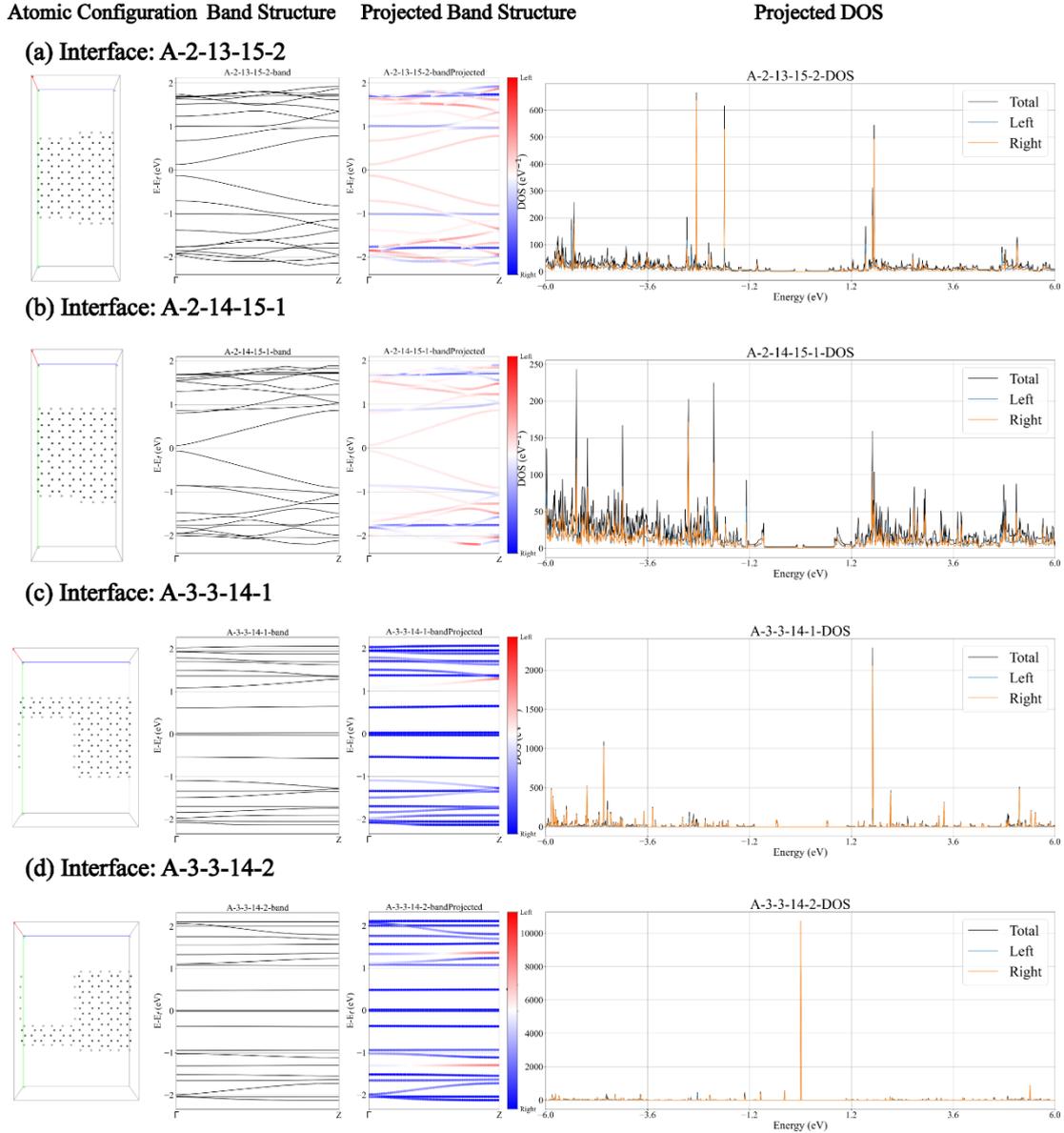

Figure 4. Some illustrations of the interface structures based on the armchair unit cell. Panels (a), (b), (c), and (d) show the interfaces A-2-13-15-2, A-2-14-15-1, A-3-3-14-1, and A-3-3-14-2, respectively. Their atomic configurations, energy band structures, projected bands, and PDOS are displayed.

We randomly select a number of interface structures in the dataset for explanatory illustration. In Figure 4, multiple AGNR segments of different widths are spliced into the interface based on the naming rule described earlier. First, the various topologies of the interfaces correspond to energy bands that exhibit bands with some curvature and fairly flat bands near the Fermi level. This will be related to the electron's ability to transport and hop. A flat band means that the energy of electrons is independent of momentum, and the effective mass of electrons tends to infinity, which corresponds to a localized distribution of electrons in real space. Then, according to the image of the projected bands, the curved bands are mainly contributed by the thinner left part and the flat bands are provided by the thicker right part. Finally, their PDOS images show in which energy intervals the number of quantum states is higher. In Figure 5, some interfaces formed by zigzag cells are shown.

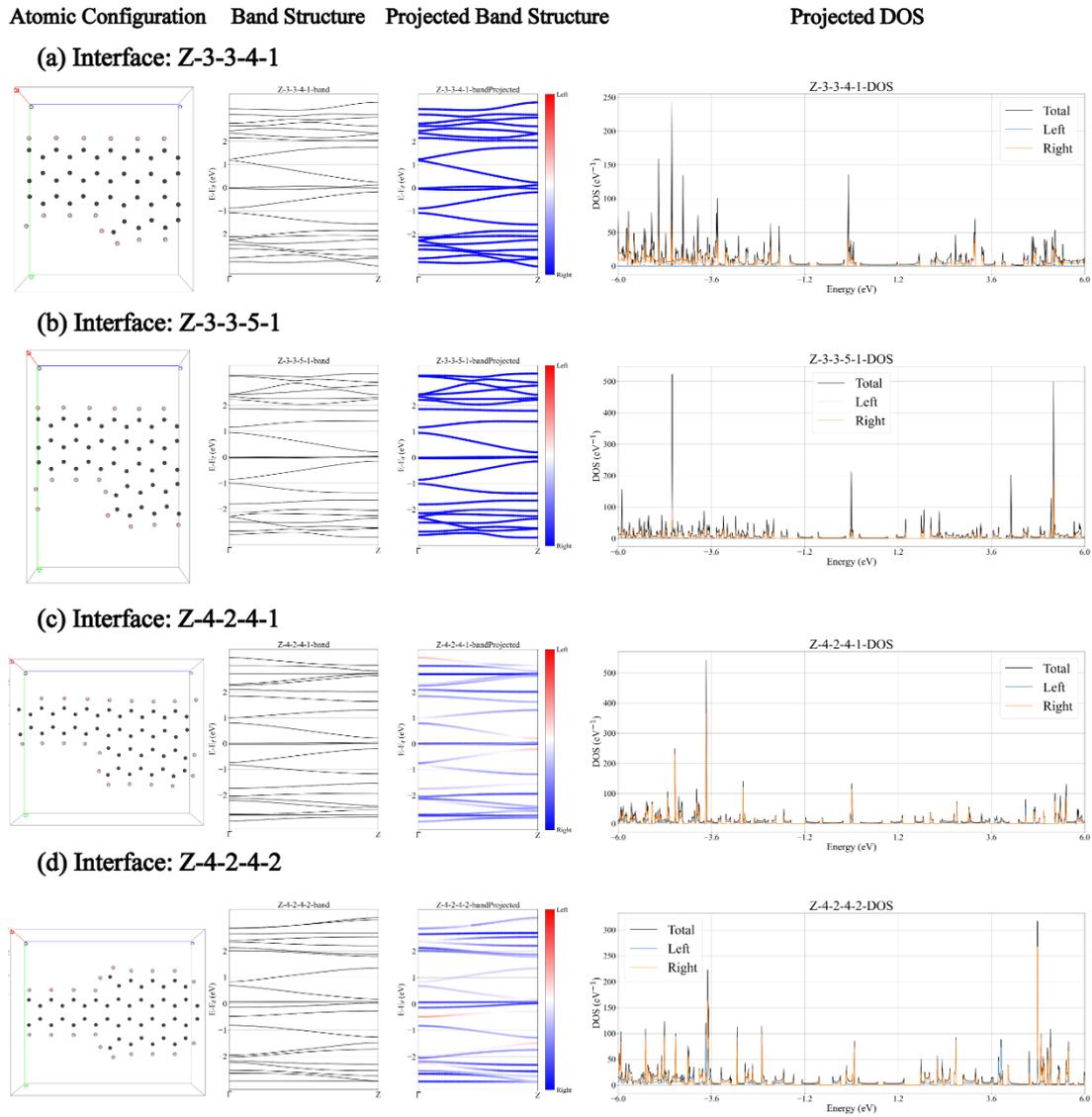

Figure 5. Some illustrations of the interface structures based on the zigzag unit cell. Panels (a), (b), (c), and (d) show the interfaces Z-3-3-4-1, Z-3-3-5-1, Z-4-2-4-1, and Z-4-2-4-2, respectively.

## Data base availability

All datasets and scripts used in this work have been made available on
https://github.com/wow2512311/Interface-Database

Cite our related work:

1. https://doi.org/10.1088/2632-2153/ad0937
   Ao Wu, et al., Graph machine learning framework for depicting wavefunction on interface. Machine Learning: Science and Technology, 2023.
2. https://doi.org/10.1103/PhysRevApplied.12.044018
   Ye, S., et al., Wave-Function Symmetry Mechanism of Quantum-Well States in Graphene Nanoribbon Heterojunctions. Physical Review Applied, 2019. 12(4): p. 044018.
3. https://doi.org/10.1103/PhysRevApplied.11.024026
   Lv, Y., et al., Interface Coupling as a Crucial Factor for Spatial Localization of Electronic States in a Heterojunction of Graphene Nanoribbons. Physical Review Applied, 2019. 11(2).

## Acknowledgements

This work is supported by the National Natural Science Foundation of China (62074116 and 81971702), and the Luojia Young Scholars Program. The numerical calculations in this paper have been done on the supercomputing system in the Supercomputing Center of Wuhan University.

## Copyright

ACTION OF CONTRACT, TORT OR OTHERWISE, ARISING FROM, OUT OF OR IN CONNECTION WITH THE SOFTWARE OR THE USE OR OTHER DEALINGS IN THE SOFTWARE.